\pdfoutput=1

\documentclass[11pt]{article}

\usepackage[]{ACL2023}

\usepackage{times}
\usepackage{latexsym}
\usepackage{amsmath}
\usepackage{booktabs}
\usepackage{multirow}
\usepackage{arydshln}
\usepackage{tikz}
\usepackage[T1]{fontenc}
\usepackage{arydshln}
\usepackage{amsmath,graphicx}
\usepackage{pgfplots}      

\usepackage[utf8]{inputenc}

\usepackage{microtype}

\usepackage{inconsolata}
\usepackage{graphicx} 
\usepackage{mdframed}

\newmdenv[
  backgroundcolor=gray!5,
  linecolor=black,
  frametitlefont=\normalfont\bfseries,
  innertopmargin=10pt,
  innerbottommargin=10pt,
  roundcorner=5pt
]{comparisonbox}

\definecolor{darkblue}{rgb}{0.0, 0.0, 1.0}

\newcommand{\myccone}{\cellcolor{gray!5}}
\newcommand{\myccfour}{\cellcolor{gray!10}}

\title{\textit{Failing Forward}: Improving Generative Error Correction for ASR \\ with Synthetic Data and Retrieval Augmentation}

\author{
    Sreyan Ghosh$^{1}$\thanks{Work done during internship at Microsoft},
    Mohammad Sadegh Rasooli$^{2}$,
     Michael Levit$^{2}$,
    Peidong Wang$^{2}$,\\
    \bf Jian Xue$^{2}$,
    \bf Dinesh Manocha$^{1}$,
    \bf  Jinyu Li$^{2}$, \\
    $^{1}$University of Maryland, College Park, USA \quad $^{2}$Microsoft, USA \\
    \texttt{Correspondence: sreyang@umd.edu} \\ 
}

\begin{document}
\maketitle
\begin{abstract}

Generative Error Correction (GEC) has emerged as a powerful post-processing method to enhance the performance of Automatic Speech Recognition (ASR) systems. However, we show that GEC models struggle to generalize beyond the specific types of errors encountered during training, limiting their ability to correct new, unseen errors at test time, particularly in out-of-domain (OOD) scenarios. This phenomenon amplifies with named entities (NEs), where, in addition to insufficient contextual information or knowledge about the NEs, novel NEs keep emerging. To address these issues, we propose \textbf{DARAG} (\textbf{Da}ta- and \textbf{R}etrieval-\textbf{A}ugmented \textbf{G}enerative Error Correction), a novel approach designed to improve GEC for ASR in in-domain (ID) and OOD scenarios. We augment the GEC training dataset with synthetic data generated by prompting LLMs and text-to-speech models, thereby simulating additional errors from which the model can learn from. For OOD scenarios, we simulate test-time errors from new domains similarly and in an unsupervised fashion. Additionally, to better handle named entities, we introduce retrieval-augmented correction by augmenting the input with entities retrieved from a database. Our approach is simple, scalable, and both domain- and language-agnostic. We experiment on multiple datasets and settings, showing that DARAG outperforms all our baselines, achieving 8\% – 30\% relative WER improvements in ID and 10\% – 33\% improvements in OOD settings.

\end{abstract}

\setlength{\abovedisplayskip}{6pt}
\setlength{\belowdisplayskip}{6pt}

\section{Introduction}

Automatic Speech Recognition (ASR) is the foundational task of converting spoken language into text. As a fundamental goal in computational language processing~\cite{jurafsky2000speech}, ASR has facilitated communication across diverse fields, including education~\cite{caballero2017asr}, healthcare~\cite{latif2020speech}, and others~\cite{den-bogaert-etal-2022-automatically}. Advances in deep learning have driven significant progress in ASR, with end-to-end models achieving impressive results on various tasks~\cite{li2022recent}. However, one of the key challenges in real-world ASR applications ~\cite{li2015robust} is handling variations in speech due to factors like background noise~\cite{chen2022noise}, speaker accents~\cite{turan-etal-2022-adapting}, and different speaking styles~\cite{syed-etal-2021-generating}. These factors lead to a significant reduction in the accuracy of ASR.

\begin{figure}
    \centering
    \includegraphics[width=\columnwidth]{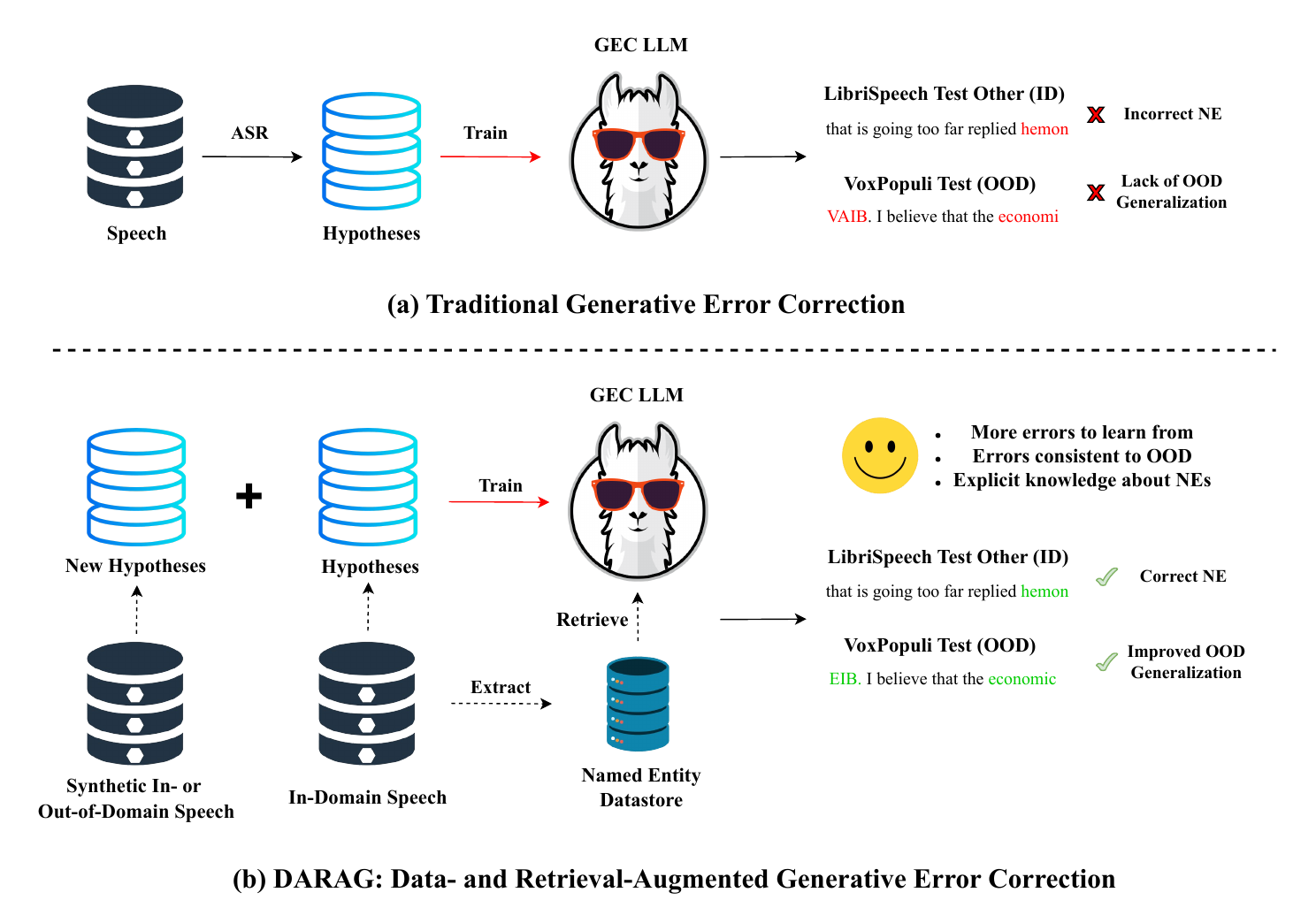}
    \vspace{-2em}
    \caption{\small Comparison of traditional GEC and DARAG. We augment the training dataset with synthetic data generated using our algorithm and named entities retrieved from a datastore to improve in-domain and out-of-domain ASR.}
    \label{fig:cartoon}
    \vspace{-1.5em}
\end{figure}

Humans demonstrate exceptional resilience to challenging speech conditions due to our inherent linguistic knowledge. Traditional ASR systems mimic this by incorporating a separate language model (LM) to rescore hypotheses during decoding~\cite{toshniwal2018comparison,kannan2018analysis}. The LM evaluates the fluency of the N-best hypotheses generated by the ASR model, and the scores are combined with the ASR's own scores in a weighted fashion. The hypothesis with the highest combined score is then selected as the final transcript. However, the rise of large language models (LLMs) with advanced reasoning capabilities has opened possibilities beyond simple rescoring. This has led to the development of Generative Error Correction (GEC)~\cite{chen2024hyporadise}, where models are trained to correct errors in the best hypothesis by leveraging information from other hypotheses, ultimately improving transcription accuracy.

Conventional Generative Error Correction (GEC) models are typically trained by pooling hypothesis-transcription pairs from various ASR systems and datasets, with the expectation that they will generalize well across diverse data at test time~\cite{chen2024hyporadise,hu2024large,ghosh23b_interspeech}. However, we identify key limitations in this approach. Previous work has primarily focused on foundational or semi-open-source models (e.g., Whisper~\cite{radford2023robust}). To explore these limitations, we conducted several single-domain, single-dataset experiments (see Table~\ref{tab:audiocaps}), training GEC models on the same datasets used to train the ASR models. We observed only minor improvements in Word Error Rate (WER) on in-domain tests and no improvements on out-of-domain (OOD) tests. Upon closer examination, we attribute these shortcomings to three main factors:
\begin{enumerate}
\vspace{-2mm}
\setlength\parskip{0em}
    \item ASR models generate too few errors on their training data for GEC models to effectively learn error correction.
    \item The GEC models are unable to generalize to the novel types of errors it sees at test time. This problem is exacerbated in OOD scenarios, where there is a significant shift in the nature of errors encountered during training versus those at test time.
    \item GEC models continue to struggle with accurately correcting novel named entities (NEs) in transcriptions. While large language models possess extensive linguistic knowledge, named entities often do not follow general language patterns. We attribute this challenge to insufficient context or a lack of knowledge about specific named entities. This problem aplifies in OOD settings.
\end{enumerate}

These observations lead us to a central hypothesis: \textit{The generalization ability of GEC models is limited by the diversity and nature of error types encountered during training.} Improving performance requires training GEC models on a broader and diverse set of errors (for richer training signals) that are consistent in its characteristics with the types the ASR model generates on the test set. To better generalize to OOD, GEC models need to be trained to correct errors that the in-domain ASR model might plausibly make on the OOD test set.

\noindent \textbf{Our Contributions.} To this end, we propose \textbf{DARAG} (\textbf{Da}ta- and \textbf{R}etrieval-\textbf{A}ugmented Generative Error Correction), a simple, scalable, and domain-agnostic approach designed to enhance GEC performance in ID and OOD scenarios. Our approach is driven by the hypothesis that GEC models perform better when trained to correct errors they are likely to encounter at test time. To achieve this, DARAG generates synthetic training data using generative models. We start by prompting an LLM with few-shot examples of domain-specific transcripts to produce synthetic transcripts. For OOD settings, DARAG uses a small set of unsupervised audio samples from the target domain, which are transcribed using the in-domain ASR model to create in-context exemplars. The synthetically generated transcripts are then used to generate synthetic speech via a text-to-speech model and voice cloning. Finally, this synthetic speech is used to generate hypothesis-transcription pairs. This process simulates errors that are specific to the target-domain vocabulary and also imitates the phonetic confusions that the ID ASR model would make in the target domain. Additionally, to improve named entity correction, we introduce a retrieval augmentation method~\cite{lewis2020retrieval}. Specifically, we extract and store all named entities from the training dataset in a datastore and retrieve the top-\textit{k} most similar entities during GEC. Our proposed method is scalable, with the datastore being easily extendable at test time to incorporate new entities as they are encountered. To summarize, our main contributions are as follows:

\begin{enumerate}
\vspace{-2mm}
\setlength\parskip{0em}
    \item We conduct a thorough investigation into the generalization limitations of LLM-based GEC, demonstrating that its performance can be improved by exposing it to diverse but consistent errors that ASR models are likely to produce at test time.

    \item To address these challenges, we propose DARAG, a novel method for enhancing GEC in both ID and OOD scenarios. DARAG augments GEC training datasets with synthetic data and decouples named entity correction from the error correction learning process through RAG. DARAG significantly outperforms traditional GEC methods, improving ASR performance by 8\%-33\%.

    \item We provide extensive ablation studies and result analysis to validate the effectiveness of DARAG across various scenarios.
    
\end{enumerate}

\section{Related Work}
\vspace{-2mm}
{\noindent \textbf{Generative Error Correction.}} Post-ASR error correction using language models (LMs) has been widely studied~\cite{ma2023can,ma2023n,zhang2023patcorrect}. Recently, large language models (LLMs) have been applied to this task, rebranded as generative error correction~\cite{hu2024large,ghosh23b_interspeech}. While LLMs excel due to their advanced language comprehension, it remains unclear which errors they effectively correct, which they miss, and how well they handle novel or unknown named entities (NEs) that they lack prior knowledge of.
\vspace{1mm}

{\noindent \textbf{Domain Generalization and Named Entity in ASR.}} Transcribing NEs is a persistent challenge for ASR models~\cite{das2022listen}. Techniques such as memorization~\cite{bekal2021remember} and biasing~\cite{jayanthi2023retrieve} have been developed to improve NE transcription. However, these methods typically focus on known NEs seen during training and struggle with unseen entities, as autoregressive models tend to memorize NEs but generalize poorly to new ones~\cite{heinzerling2020language}. Improving NE transcription using post-ASR processing or GEC has not been well explored.  ASR models often fail under distribution shifts, such as domain, accent, or dialect changes~\cite{singhal2023domain}. However, the robustness of GEC to domain shifts remains underexplored.

\section{Preliminary}
\label{subsec:prelim}

\subsection{Problem Formulation}
\label{subsec:problem_formulation}

Let $\mathcal{D}^\text{id}_\text{train} = \{(a_i,t_i), 1 \leq i \leq n\}$ represent a human-annotated, in-domain dataset containing $n$ pairs of speech and corresponding transcripts for training an ASR system ($\mathcal{D}^\text{id}_\text{train}$ is sourced from a single dataset and not pooled from multiple datasets unless otherwise mentioned). Consider $\mathcal{A}^\theta$ as an encoder-decoder ASR model trained on $\mathcal{D}_\text{gold}$. For GEC, our goal is to generate a list of N-best hypotheses $h_i$ for each instance in $\mathcal{D}^\text{id}_\text{train}$ using beam search decoding. Next, using the hypotheses and corresponding gold transcripts, denoted by $\mathcal{H}^\text{id}_\text{train} = \{(h_i, t_i), 1 \leq i \leq n\}$, we fine-tune a language model to correct the errors in the best hypothesis by leveraging cues from the other \textit{N}-1 hypotheses to directly produce an accurate transcription. During training, the true transcription $t_i$ serves as the target. At inference time, for each instance in the test set $\mathcal{D}^\text{id}_\text{test}$, we generate a list of hypotheses and prompt the fine-tuned model to output a corrected transcript. 

Our objective is to create a synthetic dataset, $\mathcal{D}^\text{id}_\text{syn} = \{(\hat{a}_i, \hat{t}_i), 1 \leq j \leq n_\text{syn}\}$, generate N-best hypotheses for each instance in it ($\hat{\mathcal{H}}^\text{id}_\text{train} = \{(\hat{h}_i, \hat{t}_i), 1 \leq j \leq n_\text{syn}\}$), and augment the original set $\mathcal{H}$ with $\hat{\mathcal{H}}$ to improve error correction on the test set $\mathcal{D}^\text{id}_\text{test}$. Alternatively, for an out-of-domain test set $\mathcal{D}^\text{ood}_\text{test}$, we assume the availability of a small train set from the same domain $\mathcal{D}^\text{ood}_\text{train} = \{(a_i, t_i), 1 \leq i \leq n_\text{small}\}$ where $n_\text{small} \ll n$ and the accompanying transcripts $t_i$ may be human-annotated or generated from $\mathcal{A}^\theta$.

\begin{table}[t]
\centering
\small
\resizebox{0.48\textwidth}{!}{
\begin{tabular}{clcc}
\toprule
\textbf{Test} & \textbf{ASR Train} & \textbf{Mismat. WER ($\downarrow$)}  & \textbf{Mat. WER ($\downarrow$)} \\
\midrule
\multirow{3}{*}{LS} 
&\myccone LS (960) (No GEC) &\myccone 4.6 &\myccone 4.6 \\ \cmidrule(l){2-4}
\multirow{3}{*}{(Clean)} & LS (960) & 4.4 & 4.4 \\
& Vox & 7.4 & \textbf{3.9} \\
& SPGI & 8.8 & 4.0 \\
\midrule
\multirow{4}{*}{Vox}
&\myccone Vox (No GEC) &\myccone 10.1 &\myccone 10.1 \\ \cmidrule(l){2-4}
& Vox & 9.4 & 9.4 \\
& LS (960) &  14.5 & \textbf{6.9} \\
& SPGI & 11.8 & 7.7 \\
\midrule
\multirow{4}{*}{SPGI}
&\myccone SPGI (No GEC) &\myccone 7.5 &\myccone 7.5 \\ \cmidrule(l){2-4}
& SPGI & 7.3 & 7.3 \\
& LS (960) & 14.2 & \textbf{4.8} \\
& Vox & 10.5 & 4.9 \\

\bottomrule
\end{tabular}
\vspace{-4mm}
}
\caption{\small Performance comparison of GEC across three different ASR benchmarks from three different domains. We evaluate and compare across two scenarios: (i) \textbf{Matched Scenario}: In this case, the hypotheses-transcription pairs for training our GEC model are derived from the Train split of the Test dataset (and not from the dataset the ASR model is trained on) (ii) \textbf{Mismatched Scenario}: In this case, the hypotheses-transcription pairs are derived from the same dataset the ASR model is trained on. We show that \textbf{(a)} For domain shifts, i.e., in cases where both the hypotheses and the ASR training dataset are from a domain different from the test, GEC leads to little to no improvement, and \textbf{(b)} For in-domain scenarios where only the hypotheses are derived from the same domain as the test, employing an ASR model trained on a different domain to derive the hypothesis boosts performance.}
\label{tab:audiocaps}
\end{table}

\subsection{What do Error Correction Models Learn to Correct?}
\label{subsec:error_correct}

Most prior work on Generative Error Correction (GEC) relies on foundational open-access ASR models, like Whisper, to generate hypotheses from various datasets and then trains GEC models on these hypotheses-transcription pairs, denoted as $\mathcal{H}^\text{id}_\text{train}$. However, because the training data used for such ASR models is often undisclosed, there is limited insight into the nature of errors present in the hypotheses and, consequently, the types of errors that the GEC models learn to correct. In this work, we aim to study error correction from a more transparent perspective. Table~\ref{tab:audiocaps} presents experiments where we train an ASR model on a single dataset (LibriSpeech (LS)~\cite{panayotov2015librispeech}, VoxPopuli~\cite{wang2021voxpopuli} (Vox), SPGIspeech~\cite{o2021spgispeech}), then derive hypotheses from either the same or a different dataset, and use these pairs to train a GEC model. Our key findings are as follows: \textbf{(i)} When GEC models are trained on a dataset in a different domain (i.e., both $\mathcal{D}^\text{id}_\text{train}$ and $\mathcal{H}^\text{id}_\text{train}$ come from a domain that is different from $\mathcal{D}^\text{id}_\text{test}$), no performance improvements are observed. We hypothesize this is due to the GEC model encountering errors at test time that differ significantly from those it saw during training. For instance, a hypotheses (HP)-transcription (GT) pair generated from the LibriSpeech train set using an ASR model trained on LibriSpeech is as follows:

\begin{comparisonbox}
\noindent \textbf{GT:} biscuits with sugar on the top preserved ginger hams brawn under glass everything in fact that makes life worth living

\noindent \textbf{HP 1:} biscuits with sugar on the top preserved ginger hams brawn under glass everything in fact that makes life worth living

\noindent \textbf{HP 2:} biscuits with sugar on the top preserved ginger hams \textcolor{red}{bran} under glass everything in fact that makes life worth living
\end{comparisonbox}

An error by the same ASR model on the VoxPopuli test set, is as follows:

\begin{comparisonbox}
\noindent \textbf{GT:} spyware allows a third party to access the same data as the user.

\noindent \textbf{HP 1:} \textcolor{red}{spygware} allows a third party to \textcolor{red}{possess} the same data as the user

\noindent \textbf{HP 2:} \textcolor{red}{spygware} allows a third party to \textcolor{red}{occupy} the same data as the user
\end{comparisonbox}

\noindent As we can see, it introduces semantic and lexical errors that are out of the domain knowledge learned during training. \textbf{(ii)} When GEC models are trained on a dataset in a similar domain (i.e., both $\mathcal{D}^\text{id}_\text{train}$ and $\mathcal{H}^\text{id}_\text{train}$ come from a domain identical to $\mathcal{D}^\text{id}_\text{test}$), improvements are minimal. We attribute this to the ASR model making fewer errors during inference, providing limited opportunities for the GEC model to learn effective corrections. For example, an ASR model trained on LibriSpeech and VoxPopuli have WERs of 2.2 and 5.1 on their respective train sets. \textbf{(iii)} To examine whether a higher error rate in hypotheses enhances GEC training, we use an ASR model trained on a different domain to generate hypotheses on our in-domain dataset $\mathcal{D}^\text{id}_\text{train}$ for GEC model training (the same ASR model is also used for test inference). Surprisingly, this setup consistently yields the most significant improvements, likely because the GEC model learns from a broader range of errors, enhancing its corrective abilities. \textit{These findings highlight (i) the need for a large and diverse set of errors and (ii)  the need for consistency in error characteristics with those that GEC models will encounter at test time.}
\vspace{-1mm}

\subsection{How Well do they fair on Named Entities?}

To assess the ability of GEC models to correct named entities (NEs), we analyze their performance in various settings. As mentioned earlier, transcribing NEs is a major challenge in ASR, particularly in knowledge-rich domains. Table~\ref{tab:ner} compares GEC performance on VoxPopuli using models trained under different conditions. For this experiment, we leverage annotated NEs from the MSNER dataset~\cite{meeus-etal-2024-msner} for VoxPopuli. Our key findings are: \textbf{(i)} GEC models struggle to correct NEs, likely due to insufficient prior knowledge or context. \textbf{(ii)} In domain-shift scenarios, where ASR or GEC models have not encountered the target NEs during training, NE transcription accuracy declines sharply. \textit{These results emphasize the importance of incorporating explicit knowledge of NEs to improve correction performance.}

\begin{table}[t]
\centering
\small
\resizebox{0.48\textwidth}{!}{
\begin{tabular}{llcc}
\toprule
\textbf{Test} & \textbf{ASR Train} & \textbf{Mismat. 
F1 ($\uparrow$)}  & \textbf{Mat. F1 ($\uparrow$)} \\
\midrule
\multirow{4}{*}{Vox}
& Vox (No GEC) & 87.8 & 87.8 \\ \cmidrule(l){2-4}
& Vox & 87.8 & 87.8 \\
& LS (960) &  80.9 & 83.2 \\
& SPGI & 81.4 & 84.0 \\
\midrule
\bottomrule
\end{tabular}
}
\caption{\small Performance comparison of GEC on VoxPopuli, an entity-rich dataset. The Matched Scenario and Mismatched Scenarios are defined as in Table~\ref{tab:audiocaps}. We show that \textbf{(a)} For domain shifts, model performance degrades significantly on NEs. \textbf{(b)} For in-domain scenarios, GEC does not prove to be effective in correcting NEs.}
\label{tab:ner}
\vspace{-3mm}
\end{table}

\begin{figure*}[t]
    \centering
\includegraphics[width=\textwidth]{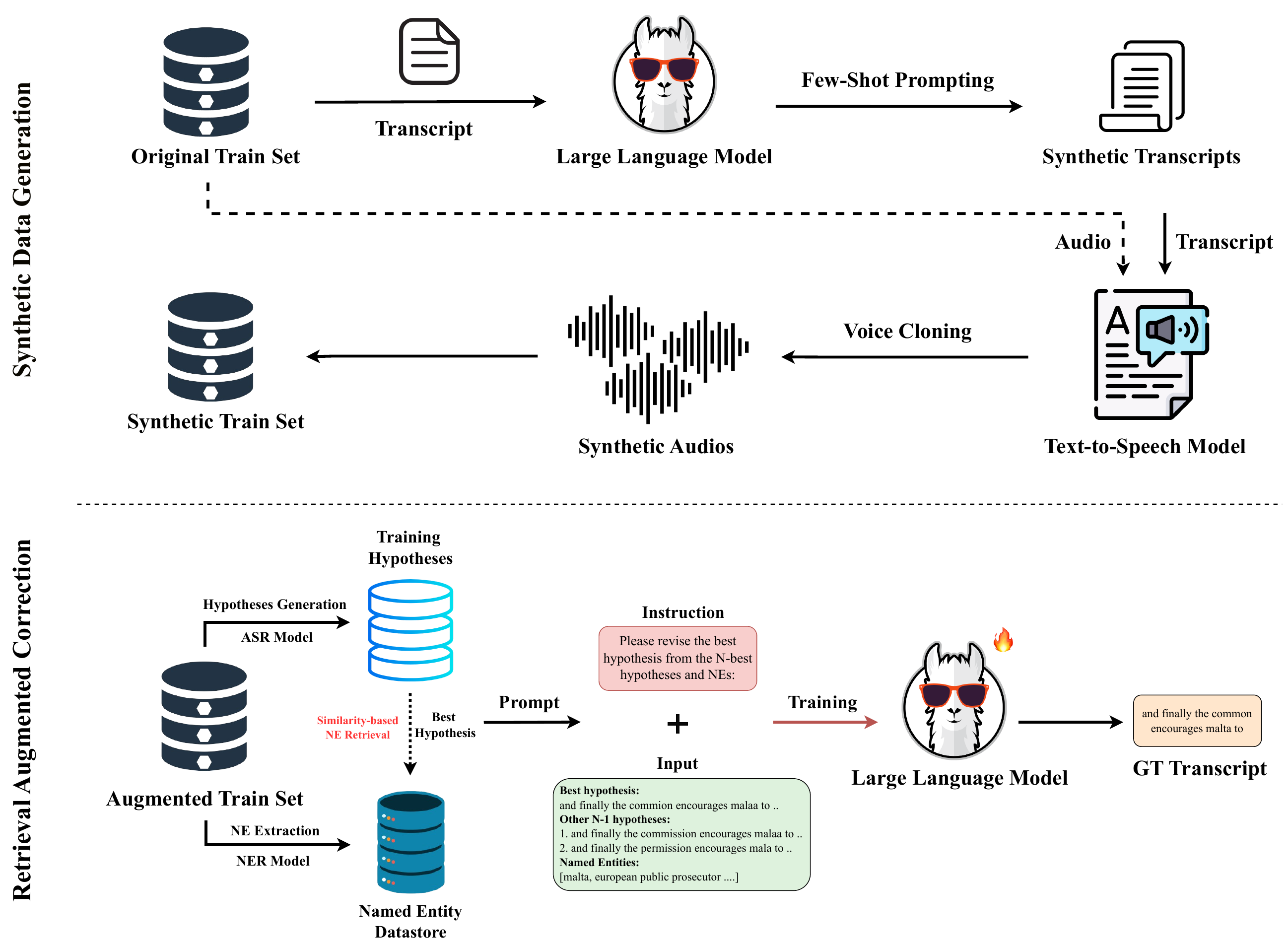}
    \caption{\small Illustration of \textbf{DARAG.} \textcircled{\raisebox{-0.9pt}{1}} We generate synthetic data with LLMs and TTS models that are then used to generate hypotheses with diverse errors consistent with the types the ASR model generates on the test set. \textcircled{\raisebox{-0.9pt}{2}} We extract the NEs and store them in a datastore. During training, for every instance, we retrieve the top-\textit{k} most similar NEs to the best hypothesis and use it to construct an instruction-response pair. \textit{Note that in OOD settings we only assume the availability of only a few unsupervised speech samples in the original train set and pseudo-transcripts for prompting are generated using the in-domain ASR model.}}
    \label{fig:main_diag}
    \vspace{-2mm}
\end{figure*}
\vspace{-1mm}
\section{Methodology}
\label{sec:methodlogy}
\vspace{-1mm}

Fig.~\ref{fig:main_diag} illustrates our proposed method. We propose two simple extensions to improve conventional GEC. First, we propose training the GEC model on additional synthetic data generated using generative models. Additionally, instead of memorizing the named entities, we propose decoupling them from the learning process with RAG. To achieve this, we first extract named entities and store them in a datastore. During training and inference, we retrieve them from the datastore and augment them to the instruction with the best hypothesis and other hypotheses. In the following subsections, we explain our methodology in detail.

\subsection{Synthetic Training Data Augmentation}
\label{subsec:synthetic}
\vspace{-1mm}

{\noindent \textbf{For In-Domain Scenarios.}} As discussed in Section~\ref{subsec:error_correct}, GEC models fail to learn effective error correction due to the low number of errors in ASR training data. We hypothesize that generating novel spoken utterances not seen during ASR training will introduce more errors that can provide rich training signals for learning error correction. Our goal is to generate spoken utterances that closely mimic the speech characteristics of speakers in the same domain, replicating the style as if spoken by similar speakers in similar contexts. These utterances can then be used to generate new hypotheses, $\hat{\mathcal{H}}^\text{id}_\text{train}$, which we augment into the original dataset $\mathcal{H}^\text{id}_\text{train}$. We achieve this through a 3-step process:

{\noindent \textbf{Step 1.}} We prompt an LLM (LLaMa-2.0-Instruct~\cite{touvron2023llama}) with in-context examples sampled from $\mathcal{D}^\text{id}_\text{train}$ to generate in-domain transcripts (prompt in Appendix~\ref{sec:prompts}).
\vspace{1mm}

{\noindent \textbf{Step 2.}} Using voice cloning via TTS,  we generate spoken utterances from the transcripts. The TTS model (Parler-TTS Mini~\cite{lacombe-etal-2024-parler-tts}) is conditioned on randomly selected utterances from $\mathcal{D}^\text{id}_\text{train}$ to replicate the domain's speech style. Steps 1 and 2 ensure the generated utterances align with the domain and produce error patterns similar to those expected at test time.
\vspace{1mm}

{\noindent \textbf{Step 3.}} We generate hypotheses for these utterances using the ASR model $\mathcal{A}^\theta$. The resulting hypotheses, $\hat{\mathcal{H}}^\text{id}_\text{train}$, are then added to $\mathcal{H}^\text{id}_\text{train}$ to improve GEC model training.

{\noindent \textbf{For Out-of-Domain Scenarios.}} In OOD settings, we follow the same steps using $\mathcal{D}^\text{ood}_\text{train}$. If annotated transcripts are unavailable, we first transcribe the utterances with the ASR model $\mathcal{A}^\theta$. Recall that in our setting $\mathcal{D}^\text{ood}_\text{train}$ only has a few utterances ($n_\text{small} \le$ 50) and is unsuitable for adaptation of $\mathcal{A}^\theta$.

\subsection{Retrieval Augmented Correction}
\label{sec:rac}

To enhance the correction of named entities (NEs), we decouple NE correction from the main GEC process and introduce a Retrieval-Augmented Correction (RAC) approach. Inspired by RAG, we retrieve the most relevant NEs during both training and inference. Our method follows three steps:

{\noindent \textbf{Step 1.}} We apply named entity recognition (NER) on all transcriptions, including those generated synthetically during the previous data augmentation step. The extracted NEs are stored in a datastore, $\mathcal{DS} = \{(s_t), 1 \leq t \leq d\}$ where $d$ is the total number of extracted NEs.

{\noindent \textbf{Step 2.}} During GEC training and inference, we use SentenceBERT~\cite{reimers2019sentence} to retrieve the top-\textit{k} NEs, $\overline{s}$, from $\mathcal{DS}$ based on their similarity to the best hypothesis. This is formally defined as:
\begin{equation}
\overline{s} = \text{top-\textit{k}}_{1 \leq t \leq d} \left(\text{sim}\left( \frac{e_i \cdot e_t}{\|e_i\| \|e_t\|} \right)\right)
\end{equation}
where $e_i$ is the SentenceBERT embedding for the best hypothesis, $e_t$ is the embedding for an NE in $\mathcal{DS}$, and $\text{sim(.)}$ is the cosine similarity between embeddings. We calculate similarity for each NE in $\mathcal{DS}$ and select the top-\textit{k} most similar NEs. This simple method proves to be extremely effective in our case as most errors in named entities belong to misspelled characters due to phonemes misrecognized by the ASR model. However, real-world datasets may contain multiple similarly spelled NEs, and retrieving all such NEs might make it difficult for error correction. We further discuss this in the limitations section.

{\noindent \textbf{Step 3.}} The retrieved NEs are then added to the input prompt during training and inference as a simple comma-separated list. We found that different prompt formats yielded similar results.

\subsection{Fine-tuning}

To train the LLM for error correction, we follow previous approaches by creating instruction-response pairs and fine-tuning our LLM on this data. The template for our instruction is as follows:

\begin{comparisonbox}
\noindent Below is the best hypothesis transcribed from a speech recognition system. Please try to revise it using the words that are only included in the other hypotheses and a list of named entities from a database, both of which will be provided to you. 

\noindent \textbf{Best-hypothesis}:

\noindent \textbf{Other-hypothesis}:

\noindent \textbf{Named-Entities}:

\end{comparisonbox}

The ground-truth transcription serves as the target for fine-tuning. Following prior work, we fine-tune only LoRA adapters~\cite{hu2021lora}.

\begin{table*}[]
\caption{\small Performance comparison (WER) of DARAG with other methods on various in-domain and out-of-domain settings (the Test is OOD w.r.t. the Train). We assume all 5 datasets are from different domains. We also report the absolute improvements w.r.t. to the ASR-only Baseline. DARAG outperforms other methods by 8\%–30\% in in-domain and 10\%–33\% in OOD settings.}
\label{tab:my-table}
\resizebox{2.1\columnwidth}{!}{
\begin{tabular}{ccccc|cccc|c|ccc}
\toprule
\multirow{2}{*}{Test} & \multicolumn{1}{l}{} & \multirow{2}{*}{Train} & \multirow{2}{*}{Baseline} & \multirow{2}{*}{Synth. Adapt.} & \multirow{2}{*}{+LM$_{rank}$} & \multirow{2}{*}{+Enhance} & \multirow{2}{*}{+GER} & \multirow{2}{*}{+RobustGER} & +DARAG & w/o RAC & w/o Aug. & only Synth. \\
\multicolumn{1}{l}{} & \multicolumn{1}{l}{} & \multicolumn{1}{l}{} &  &  &  &  &  &  & \textit{(ours)} & \textit{(ours)} & \textit{(ours)} & \textit{(ours)} \\
\midrule
& \textit{In-Domain} & LS & 4.6 & 4.6$_{\textcolor{red}{+0\%}}$& 4.4$_{\textcolor{teal}{-4.3\%}}$ & 4.5$_{\textcolor{teal}{-2.2\%}}$ & 4.4$_{\textcolor{teal}{-4.3\%}}$ & 4.4$_{\textcolor{teal}{-4.3\%}}$ & \cellcolor{gray!5}\textbf{4.0}$_{\textcolor{teal}{-13.0\%}}$& \cellcolor{gray!5}4.2$_{\textcolor{teal}{-8.7\%}}$& \cellcolor{gray!5}\underline{4.1}$_{\textcolor{teal}{-10.9\%}}$& \cellcolor{gray!5}4.6$_{\textcolor{red}{+0.0\%}}$\\
\cmidrule(l){2-13}
& & Vox & 8.2 & 8.8$_{\textcolor{red}{+7.3\%}}$& 8.1$_{\textcolor{teal}{-1.2\%}}$ & 8.3$_{\textcolor{teal}{-1.2\%}}$  & 7.4$_{\textcolor{teal}{-9.8\%}}$ & 7.4$_{\textcolor{teal}{-9.8\%}}$ & \cellcolor{gray!5}\underline{6.1}$_{\textcolor{teal}{-25.6\%}}$& \cellcolor{gray!5}\textbf{5.9}$_{\textcolor{teal}{-28.0\%}}$& \cellcolor{gray!5}6.8$_{\textcolor{teal}{-17.1\%}}$&  \cellcolor{gray!5}\\
& & SPGI & 8.9 & 9.0$_{\textcolor{red}{+1.1\%}}$& 8.8$_{\textcolor{teal}{-1.1\%}}$ & 8.8$_{\textcolor{teal}{-1.1\%}}$ & 8.8$_{\textcolor{teal}{-1.1\%}}$ & 8.6$_{\textcolor{teal}{-3.4\%}}$ & \cellcolor{gray!5}\underline{8.0}$_{\textcolor{teal}{-10.1\%}}$& \cellcolor{gray!5}\textbf{7.8}$_{\textcolor{teal}{-12.4\%}}$& \cellcolor{gray!5}\underline{8.0}$_{\textcolor{teal}{-10.1\%}}$& \cellcolor{gray!5}16.8$_{\textcolor{red}{+44.8\%}}$\\
\multirow{-4}{*}{LS (Clean)} & \multirow{-3}{*}{\textit{Out-of-Domain}} & TED & 11.6 & 11.5$_{\textcolor{teal}{-0.9\%}}$& 11.1$_{\textcolor{teal}{-4.3\%}}$ & 11.4$_{\textcolor{teal}{-4.3\%}}$ & 11.3$_{\textcolor{teal}{-2.6\%}}$ & 11.3$_{\textcolor{teal}{-2.6\%}}$ & \cellcolor{gray!5}\underline{10.2}$_{\textcolor{teal}{-12.1\%}}$& \cellcolor{gray!5}\textbf{9.9}$_{\textcolor{teal}{-14.7\%}}$& \cellcolor{gray!5}10.9$_{\textcolor{teal}{-6.0\%}}$&  \cellcolor{gray!5}\\
\midrule
& \textit{In-Domain} & LS & 8.4 & 8.3$_{\textcolor{teal}{-1.2\%}}$& 7.7$_{\textcolor{teal}{-8.3\%}}$ & 7.2$_{\textcolor{teal}{-14.3\%}}$& 7.2$_{\textcolor{teal}{-14.3\%}}$ & 6.9$_{\textcolor{teal}{-17.9\%}}$ & \cellcolor{gray!5}\textbf{6.4}$_{\textcolor{teal}{-23.8\%}}$& \cellcolor{gray!5}7.0$_{\textcolor{teal}{-16.7\%}}$& \cellcolor{gray!5}\underline{6.6}$_{\textcolor{teal}{-21.4\%}}$& \cellcolor{gray!5}8.0$_{\textcolor{teal}{-4.8\%}}$\\
\cmidrule(l){2-13}
& & Vox & 13.7 & 14.0$_{\textcolor{red}{+2.2\%}}$& 13.5$_{\textcolor{teal}{-1.5\%}}$ & 13.2$_{\textcolor{teal}{-1.5\%}}$ & 13.5$_{\textcolor{teal}{-1.5\%}}$ & 13.5$_{\textcolor{teal}{-1.5\%}}$ & \cellcolor{gray!5}\textbf{11.9}$_{\textcolor{teal}{-13.1\%}}$& \cellcolor{gray!5}\underline{13.0}$_{\textcolor{teal}{-5.1\%}}$& \cellcolor{gray!5}\underline{13.0}$_{\textcolor{teal}{-5.1\%}}$&  \cellcolor{gray!5}\\
& & SPGI & 14.2 & 15.5$_{\textcolor{red}{+9.2\%}}$& 14.0$_{\textcolor{teal}{-1.4\%}}$ & 13.5$_{\textcolor{teal}{-1.4\%}}$ & 13.8$_{\textcolor{teal}{-2.8\%}}$ & 13.8$_{\textcolor{teal}{-2.8\%}}$ & \cellcolor{gray!5}\textbf{12.6}$_{\textcolor{teal}{-11.3\%}}$& \cellcolor{gray!5}\underline{13.4}$_{\textcolor{teal}{-5.6\%}}$& \cellcolor{gray!5}\underline{13.4}$_{\textcolor{teal}{-5.6\%}}$& \cellcolor{gray!5}19.2$_{\textcolor{red}{+7.2\%}}$\\
\multirow{-4}{*}{LS (Other)} & \multirow{-3}{*}{\textit{Out-of-Domain}} & TED & 17.9 & 18.6$_{\textcolor{red}{+3.9\%}}$& 17.9$_{\textcolor{red}{+0.0\%}}$ & 17.5$_{\textcolor{red}{+0.0\%}}$ & 17.4$_{\textcolor{teal}{-2.8\%}}$ & 17.4$_{\textcolor{teal}{-2.8\%}}$ & \cellcolor{gray!5}\textbf{15.3}$_{\textcolor{teal}{-14.5\%}}$& \cellcolor{gray!5}\underline{15.8}$_{\textcolor{teal}{-11.7\%}}$& \cellcolor{gray!5}16.0$_{\textcolor{teal}{-10.6\%}}$&  \cellcolor{gray!5}\\
\midrule
& \textit{In-Domain} & Vox & 10.1 &  9.9$_{\textcolor{teal}{-2.0\%}}$& 9.5$_{\textcolor{teal}{-5.9\%}}$ & 9.9$_{\textcolor{teal}{-2.0\%}}$ & 9.4$_{\textcolor{teal}{-6.9\%}}$ & 9.4$_{\textcolor{teal}{-6.9\%}}$ & \cellcolor{gray!5}\textbf{8.6}$_{\textcolor{teal}{-14.9\%}}$& \cellcolor{gray!5}9.4$_{\textcolor{teal}{-6.9\%}}$& \cellcolor{gray!5}\underline{8.9}$_{\textcolor{teal}{-11.9\%}}$& \cellcolor{gray!5}9.5$_{\textcolor{teal}{-5.9\%}}$\\
\cmidrule(l){2-13}
& & LS & 14.9  & 15.2$_{\textcolor{red}{+2.0\%}}$& 14.9$_{\textcolor{red}{+0.0\%}}$ & 14.9$_{\textcolor{red}{+0.0\%}}$ & 14.5$_{\textcolor{teal}{-2.7\%}}$ & 14.5$_{\textcolor{teal}{-2.7\%}}$ & \cellcolor{gray!5}\underline{10.0}$_{\textcolor{teal}{-32.9\%}}$& \cellcolor{gray!5}\textbf{9.8}$_{\textcolor{teal}{-34.2\%}}$& \cellcolor{gray!5}12.1$_{\textcolor{teal}{-18.8\%}}$&  \cellcolor{gray!5}\\
& & SPGI & 11.8  & 13.4$_{\textcolor{red}{+13.6\%}}$& 11.4$_{\textcolor{teal}{-3.4\%}}$ & 11.8$_{\textcolor{teal}{-3.4\%}}$ & 11.8$_{\textcolor{red}{+0.0\%}}$ & 11.6$_{\textcolor{teal}{-1.7\%}}$ & \cellcolor{gray!5}\textbf{8.1}$_{\textcolor{teal}{-31.4\%}}$& \cellcolor{gray!5}\underline{8.4}$_{\textcolor{teal}{-28.8\%}}$& \cellcolor{gray!5}10.3$_{\textcolor{teal}{-12.7\%}}$& \cellcolor{gray!5}19.8$_{\textcolor{red}{+16.5\%}}$\\
\multirow{-4}{*}{Vox} & \multirow{-3}{*}{\textit{Out-of-Domain}} & TED & 17.0  & 18.6$_{\textcolor{red}{+9.4\%}}$& 17.0$_{\textcolor{red}{+0.0\%}}$ & 17.2$_{\textcolor{red}{+0.0\%}}$ & 17.3$_{\textcolor{red}{+1.8\%}}$& 17.3$_{\textcolor{red}{+1.8\%}}$& \cellcolor{gray!5}\textbf{14.4}$_{\textcolor{teal}{-15.3\%}}$& \cellcolor{gray!5}\underline{14.7}$_{\textcolor{teal}{-13.5\%}}$& \cellcolor{gray!5}15.9$_{\textcolor{teal}{-6.5\%}}$&  \cellcolor{gray!5}\\
\midrule
& \textit{In-Domain} & TED & 6.7  & 6.5$_{\textcolor{teal}{-3.0\%}}$& 6.6$_{\textcolor{teal}{-1.5\%}}$ & 6.7$_{\textcolor{red}{+0.0\%}}$ & 6.6$_{\textcolor{teal}{-1.5\%}}$ & 6.8$_{\textcolor{red}{+1.5\%}}$& \cellcolor{gray!5}\textbf{6.2}$_{\textcolor{teal}{-7.5\%}}$& \cellcolor{gray!5}\underline{6.3}$_{\textcolor{teal}{-6.0\%}}$& \cellcolor{gray!5}6.6$_{\textcolor{teal}{-1.5\%}}$& \cellcolor{gray!5}7.0$_{\textcolor{red}{+4.5\%}}$\\
\cmidrule(l){2-13}
& & SPGI & 10.4  & 10.0$_{\textcolor{teal}{-3.8\%}}$& 10.2$_{\textcolor{teal}{-1.9\%}}$ & 10.4$_{\textcolor{teal}{-1.9\%}}$ & 10.8$_{\textcolor{red}{+3.8\%}}$& 10.8$_{\textcolor{red}{+3.8\%}}$& \cellcolor{gray!5}\underline{8.8}$_{\textcolor{teal}{-15.4\%}}$& \cellcolor{gray!5}\textbf{8.1}$_{\textcolor{teal}{-22.1\%}}$& \cellcolor{gray!5}10.1$_{\textcolor{teal}{-2.9\%}}$&  \cellcolor{gray!5}\\
& & LS & 9.1  & 9.0$_{\textcolor{teal}{-1.1\%}}$& 8.8$_{\textcolor{teal}{-3.3\%}}$ & 9.1$_{\textcolor{teal}{-3.3\%}}$ & \underline{8.5}$_{\textcolor{teal}{-6.6\%}}$ & \underline{8.5}$_{\textcolor{teal}{-6.6\%}}$ & \cellcolor{gray!5}\textbf{8.2}$_{\textcolor{teal}{-9.9\%}}$& \cellcolor{gray!5}8.7$_{\textcolor{teal}{-4.4\%}}$& \cellcolor{gray!5}\textbf{8.2}$_{\textcolor{teal}{-9.9\%}}$& \cellcolor{gray!5}15.8$_{\textcolor{red}{+51.9\%}}$\\
\multirow{-4}{*}{TED} & \multirow{-3}{*}{\textit{Out-of-Domain}} & Vox & 9.9  & 10.8$_{\textcolor{red}{+9.1\%}}$& 9.9$_{\textcolor{red}{+0.0\%}}$ & 9.9$_{\textcolor{red}{+0.0\%}}$ & 10.2$_{\textcolor{red}{+3.0\%}}$& 10.2$_{\textcolor{red}{+3.0\%}}$& \cellcolor{gray!5}\underline{9.0}$_{\textcolor{teal}{-9.1\%}}$& \cellcolor{gray!5}\textbf{8.9}$_{\textcolor{teal}{-10.1\%}}$& \cellcolor{gray!5}10.1$_{\textcolor{red}{+2.0\%}}$&  \cellcolor{gray!5}\\
\midrule
& \textit{In-Domain} & Giga & 11.5  & 14.8$_{\textcolor{red}{+28.7\%}}$& 10.8$_{\textcolor{teal}{-6.1\%}}$ & 10.6$_{\textcolor{teal}{-7.8\%}}$ & 11.0$_{\textcolor{teal}{-4.3\%}}$ & 10.6$_{\textcolor{teal}{-7.8\%}}$ & \cellcolor{gray!5}\textbf{9.1}$_{\textcolor{teal}{-20.9\%}}$& \cellcolor{gray!5}10.2$_{\textcolor{teal}{-11.3\%}}$& \cellcolor{gray!5}\underline{9.5}$_{\textcolor{teal}{-17.4\%}}$& \cellcolor{gray!5}11.0$_{\textcolor{teal}{-4.3\%}}$\\
\cmidrule(l){2-13}
& & TED & 22.7  & 24.3$_{\textcolor{red}{+7.0\%}}$& 21.5$_{\textcolor{teal}{-5.3\%}}$ & 21.8$_{\textcolor{teal}{-5.3\%}}$ & 22.3$_{\textcolor{teal}{-1.8\%}}$ & 22.3$_{\textcolor{teal}{-1.8\%}}$ & \cellcolor{gray!5}\textbf{18.5}$_{\textcolor{teal}{-18.5\%}}$& \cellcolor{gray!5}\textbf{18.5}$_{\textcolor{teal}{-18.5\%}}$& \cellcolor{gray!5}\underline{21.3}$_{\textcolor{teal}{-6.2\%}}$&  \cellcolor{gray!5}\\
& & LS & 18.0  & 23.4$_{\textcolor{red}{+30.0\%}}$& 17.7$_{\textcolor{teal}{-1.7\%}}$ & 17.5$_{\textcolor{teal}{-1.7\%}}$ & 17.8$_{\textcolor{teal}{-1.1\%}}$ & 17.8$_{\textcolor{teal}{-1.1\%}}$ & \cellcolor{gray!5}\underline{14.7}$_{\textcolor{teal}{-18.3\%}}$& \cellcolor{gray!5}\textbf{14.4}$_{\textcolor{teal}{-20.0\%}}$& \cellcolor{gray!5}16.9$_{\textcolor{teal}{-6.1\%}}$& \cellcolor{gray!5}26.2$_{\textcolor{red}{+15.4\%}}$\\
\multirow{-4}{*}{Giga} & \multirow{-3}{*}{\textit{Out-of-Domain}} & Vox & 16.3  & 20.2$_{\textcolor{red}{+23.9\%}}$& 16.2$_{\textcolor{teal}{-0.6\%}}$ & 16.2$_{\textcolor{teal}{-0.6\%}}$ & 16.6$_{\textcolor{red}{+1.8\%}}$& 16.6$_{\textcolor{red}{+1.8\%}}$& \cellcolor{gray!5}\textbf{14.5}$_{\textcolor{teal}{-11.0\%}}$& \cellcolor{gray!5}\underline{15.0}$_{\textcolor{teal}{-8.0\%}}$& \cellcolor{gray!5}16.4$_{\textcolor{red}{+0.6\%}}$&  \cellcolor{gray!5}\\
\midrule
& \textit{In-Domain} & SPGI & 7.5  & 11.0$_{\textcolor{red}{+46.7\%}}$& 7.1$_{\textcolor{teal}{-5.3\%}}$ & 7.4$_{\textcolor{teal}{-1.3\%}}$ & 7.3$_{\textcolor{teal}{-2.7\%}}$ & 7.4$_{\textcolor{teal}{-1.3\%}}$ & \cellcolor{gray!5}\textbf{5.2}$_{\textcolor{teal}{-30.7\%}}$& \cellcolor{gray!5}\underline{6.0}$_{\textcolor{teal}{-20.0\%}}$& \cellcolor{gray!5}6.4$_{\textcolor{teal}{-14.7\%}}$& \cellcolor{gray!5}7.6$_{\textcolor{red}{+1.3\%}}$\\
\cmidrule(l){2-13}
& & TED & 17.7  & 24.6$_{\textcolor{red}{+39.0\%}}$& 17.4$_{\textcolor{teal}{-1.7\%}}$ & 17.6$_{\textcolor{teal}{-1.7\%}}$ & 17.7$_{\textcolor{red}{+0.0\%}}$ & 17.7$_{\textcolor{red}{+0.0\%}}$ & \cellcolor{gray!5}\textbf{13.9}$_{\textcolor{teal}{-21.5\%}}$& \cellcolor{gray!5}\underline{14.4}$_{\textcolor{teal}{-18.6\%}}$& \cellcolor{gray!5}17.0$_{\textcolor{teal}{-4.0\%}}$&  \cellcolor{gray!5}\\
& & LS & 14.4  & 18.1$_{\textcolor{red}{+25.7\%}}$& 14.4$_{\textcolor{red}{+0.0\%}}$ & 14.4$_{\textcolor{red}{+0.0\%}}$ & 14.2$_{\textcolor{teal}{-1.4\%}}$ & 14.2$_{\textcolor{teal}{-1.4\%}}$ & \cellcolor{gray!5}\underline{12.0}$_{\textcolor{teal}{-16.7\%}}$& \cellcolor{gray!5}\textbf{11.6}$_{\textcolor{teal}{-19.4\%}}$& \cellcolor{gray!5}13.4$_{\textcolor{teal}{-6.9\%}}$& \cellcolor{gray!5}24.9$_{\textcolor{red}{+40.7\%}}$\\
\multirow{-4}{*}{SPGI} & \multirow{-3}{*}{\textit{Out-of-Domain}} & Vox & 11.3  & 14.7$_{\textcolor{red}{+30.1\%}}$& 10.9$_{\textcolor{teal}{-3.5\%}}$ & 11.0$_{\textcolor{teal}{-3.5\%}}$ & 10.5$_{\textcolor{teal}{-7.1\%}}$ & 10.4$_{\textcolor{teal}{-7.9\%}}$ & \cellcolor{gray!5}\underline{8.2}$_{\textcolor{teal}{-27.4\%}}$& \cellcolor{gray!5}\textbf{8.0}$_{\textcolor{teal}{-29.2\%}}$& \cellcolor{gray!5}10.1$_{\textcolor{teal}{-10.6\%}}$&  \cellcolor{gray!5}\\
\bottomrule
\end{tabular}}
\vspace{-2mm}
\end{table*}

\section{Experimental Setup}
\vspace{-2.5mm}

{\noindent \textbf{Models and Hyper-Parameters.}} For our ASR model, we employ an encoder-decoder model with a 12-layer transformer-based encoder and a 6-layer conformer-based decoder. We train all datasets for 100 epochs with Adam optimizer, a learning rate of 1e-3, and an effective batch size of 128. For learning GEC, we train the LLM for 10 epochs with Adam optimizer, a learning rate of 5e-5, and an effective batch size of 32. We used a standard rank of 8, and we did not find a substantial change in performance by decreasing or increasing it. We generate $n_\text{syn}$ = $n$ or as many synthetic augmentations as the size of the original training set. For top-\textit{k} NE retrieval, we set \textit{k}=5. For N-best hypotheses, we set N=5. For OOD, we set $n_\text{small}$=100. All results are averaged over 3 runs for 3 random seeds.

{\noindent \textbf{Datasets.}} We evaluate DARAG on 5 benchmark ASR datasets, including LibriSpeech-960 (LS), SPGISpeech (SPGI), VoxPopuli\textsubscript{en}(Vox), Gigaspeech~\cite{GigaSpeech2021} (Giga) and TED-LIUM~\cite{rousseau2012ted} (TED). \textit{We acknowledge that for OOD evaluation, prior works use different and varied settings. However, we want to emphasize that OOD adaptation or evaluation is not our main focus; rather, only to show DARAG improves performance in typical OOD settings.}

{\noindent \textbf{Comparison Methods and Ablations.}} For comparison with DARAG, we employ (i) Baseline -- Only ASR, and we perform no post-processing. (ii) Synth. Adap. -- For ID, we add the synthetic data to the original ASR training data. For OOD, we do adapter-based continual fine-tuning of the ASR model (full-fine-tuning gave us worse performance) (iii) GER~\cite{chada-etal-2021-error} -- This can be considered as DARAG without data aur retrieval augmentation (iv) RobustGER~\cite{radford2023robust} (v) LM$_{rank}$ -- We use the same LLM (continually fine-tuned on the text from training and synthetic dataset) as GER for re-scoring the $N$-best hypotheses and finally take the hypothesis with the best score averaged across the LLM and ASR model scores. (vi) Enhance -- we also employ a speech enhancement front-end, a HiFi-GAN~\cite{su2020hifi}, to denoise the noisy speech before passing it to the ASR model. For ablations, we employ (i) w/o RAC: DARAG without retrieval augmented correction. (ii) w/o Aug.: DARAG without synthetic data augmentation but only retrieval augmentation based error correction. (iii) only Synth.: The GEC model is only trained on hypotheses-transcription pairs from only the synthetically generated data.
\vspace{-2mm}
\section{Results and Analysis}

{\noindent \textbf{Main Results.}} Table~\ref{tab:my-table} presents our main results, comparing performance across five datasets in both ID and OOD scenarios. In the ID setting, the training and test sets come from the same dataset, whereas in the OOD setting, the training set is sourced from a different dataset, making the test set OOD for both the ASR and GEC models. For the OOD experiments, we randomly selected three datasets for training without any particular preference. Additionally, we did not assume the availability of ground-truth transcripts in $\mathcal{D}^\text{ood}_\text{train}$ and instead used our ASR model to generate transcripts. Unlike previous experiments, we did not assume a separate dataset for ASR training; both the ASR model and hypotheses were generated from the same training data. Our key findings can be summarized as follows: \textbf{(i)} DARAG substantially improves ASR performance for both ID (8\%-30\%) and OOD (10\%-33\%) settings. \textbf{(ii)} In ID settings, both RAC and synthetic augmentation prove essential, as ablating either component leads to a decline in performance. \textbf{(iii)} In OOD settings, augmentation is more beneficial than RAC, likely because most NEs in the datastore do not match the NEs encountered during testing. \textbf{(iv)} DARAG proves to be a better way to utilize synthetic data for improving ASR as adaptation with synthetic data leads to performance decrease over baseline.\textbf{(v)} In some OOD cases, removing RAC actually improves performance, which we attribute to mismatched OOD NEs causing the GEC model to incorrectly adjust certain NEs. \textbf{(vi)} Relying solely on synthetic data is not effective for OOD scenarios, consistent with prior research indicating that human-annotated data remains crucial for optimal performance~\cite{ghosh2024synthio}. \textit{In Appendix~\ref{sec:llm_copying} we demonstrate that DARAG does not simply replicate the original training data as a result of LLM memorization.}

\subsection{Does Retrieval Augmentation Improve Transcription of Named Entities?}

Table~\ref{tab:ner_end} presents a comparison of F1-micro scores for DARAG and various baselines in both ID and OOD settings. The results reveal several key insights: (i) DARAG consistently outperforms the baseline and conventional GEC approaches, with particularly large gains in OOD scenarios, demonstrating its robustness to domain shifts. (ii) Incorporating a datastore containing NEs from the in-domain dataset significantly improves OOD performance, in some cases matching the results of GEC models trained on ID datasets. This highlights the effectiveness of retrieval-augmented correction in enhancing ASR performance, including practical applications like meeting applications, where a datastore can be constructed with a list of relevant NEs and not necessarily included during training. (iii) Augmenting the datastore with synthetically generated NEs also shows promise in boosting DARAG's performance, indicating the potential to dynamically add emerging NEs to the datastore. This approach reduces the reliance on continual fine-tuning for ASR adaptation, which is typically required in other methods~\cite{das2022listen}.

\begin{table}[t]
\centering
\small
\resizebox{0.48\textwidth}{!}{
\begin{tabular}{clcc}
\toprule
\textbf{Test} & \textbf{Method} & \textbf{OOD F1 ($\uparrow$)} & \textbf{ID F1 ($\uparrow$)} \\
\midrule
\multirow{6}{*}{Vox}
&\myccfour Baseline &\myccfour 79.5   &\myccfour 87.8\\\cmidrule{2-4}
& \hspace{0.25cm}+GEC & 80.9  & 87.8\\ 
& \hspace{0.25cm}+DARAG & 82.3 & \underline{90.0} \\ 
& \hspace{0.55cm}+synth. NE & 82.8  & \textbf{92.3}\\
& \hspace{0.25cm}+DARAG w/ ID NE & \underline{89.9} & - \\ 
& \hspace{0.55cm}+synth. NE & \textbf{90.7} & - \\\cmidrule(l){1-4}
\multirow{6}{*}{LS}
&\myccfour Baseline &\myccfour 82.5 &\myccfour 93.2\\\cmidrule{2-4} 
\multirow{6}{*}{(Other)}& \hspace{0.25cm}+GEC & 82.0  & 93.5\\ 
& \hspace{0.25cm}+DARAG & 83.1 & \underline{96.0}\\
& \hspace{0.55cm}+synth. NE & 84.9  & \textbf{96.4}\\
& \hspace{0.25cm}+DARAG w/ ID NE & \underline{93.1} & -\\
& \hspace{0.55cm}+synth. NE & \textbf{93.4}  & -\\
\bottomrule
\end{tabular}
}
\vspace{-3mm}
\caption{\small Performance comparison of DARAG with other methods on the NE transcription. For ID, we employ the train set of the dataset as the test. For OOD, we employ LS for Vox and Vox for LS. w/ ID NE refers to DARAG, where the NE datastore is from the ID train set. w/ synth NE refers to additional synthetic NEs we add to the NE datastore.}
\label{tab:ner_end}
\vspace{-4mm}
\end{table}

\subsection{DARAG for Source-Free UDA}
Most Unsupervised Domain Adaptation (UDA) methods for ASR assume the presence of the entire unlabeled dataset from the target domain~\cite{hu2024self}. On the other hand, DARAG assumes the presence of only few unlabeled instances. Fig.~\ref{fig:abs_improvement} shows DARAG proves to be effective for extreme low-resource UDA and outperforms STAR and continual fine-tuning with pseudo-labeling.

\subsection{Real Data Outperforms Synthetic}

Table~\ref{tab:real_vs} shows a comparison between DARAG and various baseline configurations where the synthetic dataset is replaced with the original training set of the target domain. The results clearly demonstrate that using real training data for generating GEC hypotheses significantly boosts performance, often surpassing complete ID settings. We attribute this improvement to two main factors: (i) the ASR model produces more errors on the GEC training dataset due to domain mismatch, providing richer training signals, and (ii) the datastore is enriched with real NEs from the original training set, offering more accurate context for corrections.
\vspace{2mm}

{\noindent \textbf{Extra Results.}} We present extra results in the Appendix, including key hyper-parameter tuning results, importance of the voice cloning module and other results. Additionally, we provide examples of generated augmentations in Table~\ref{tab:example_transcripts} and DARAG corrections in Table~\ref{tab:corrections}.

\begin{table}[t]
\centering
\small
\resizebox{0.48\textwidth}{!}{
\begin{tabular}{llccc}
\toprule
\textbf{Test} & \textbf{Method} & \textbf{ASR Train} & \textbf{GEC Train} & \textbf{WER ($\downarrow$)} \\
\midrule
\multirow{10}{*}{Vox}
&\myccfour Baseline &\myccfour Vox &\myccfour-  &\myccfour 10.1 \\
&\myccfour \hspace{0.25cm}+DARAG  &\myccfour Vox &\myccfour Vox &\myccfour 8.6 \\\cmidrule{2-5}
& Baseline & LS & -  & 14.9 \\
& Baseline & LS + Vox & -  & 10.3 \\
& \hspace{0.25cm}+DARAG  & LS & LS & 10.0 \\
& \hspace{0.25cm}+DARAG & LS & Vox & \textbf{6.9} \\\cmidrule{2-5} 
& Baseline & TED & - & 17.0 \\
& Baseline & TED + Vox & - & 10.0 \\
& \hspace{0.25cm}+DARAG  & TED & TED & 14.4 \\ 
& \hspace{0.25cm}+DARAG & TED & Vox & \underline{7.5} \\\midrule
\multirow{10}{*}{SPGI}
&\myccfour Baseline &\myccfour SPGI &\myccfour -  &\myccfour 7.5 \\
&\myccfour \hspace{0.25cm}+DARAG &\myccfour SPGI &\myccfour SPGI &\myccfour 5.2 \\\cmidrule{2-5}
& Baseline & LS & -  & 13.3 \\
& Baseline & LS + SPGI & -  & 7.7 \\
& \hspace{0.25cm}+DARAG & LS & LS & 12.0 \\ 
& \hspace{0.25cm}+DARAG & LS & SPGI & \textbf{4.8} \\\cmidrule{2-5}
& Baseline & TED & - & 17.7 \\
& Baseline & TED + SPGI & - & 7.9 \\
& \hspace{0.25cm}+DARAG  & TED & TED & 13.9 \\ 
& \hspace{0.25cm}+DARAG & TED & SPGI & \underline{5.0} \\
\bottomrule
\end{tabular}
}
\vspace{-2mm}
\caption{\small Performance comparison of DARAG in OOD settings with the baseline. We replace the generated augmentations with the original target domain training dataset (and do not generate extra augmentations). Training on hypotheses from the target domain train set leads to superior performance.}
\label{tab:real_vs}
\vspace{-2mm}
\end{table}

\definecolor{Method1Color}{rgb}{0.1, 0.2, 0.7}
\definecolor{Method2Color}{rgb}{0.7, 0.2, 0.1}
\definecolor{Method3Color}{rgb}{0.2, 0.7, 0.2}

\pgfplotsset{compat=1.18} 
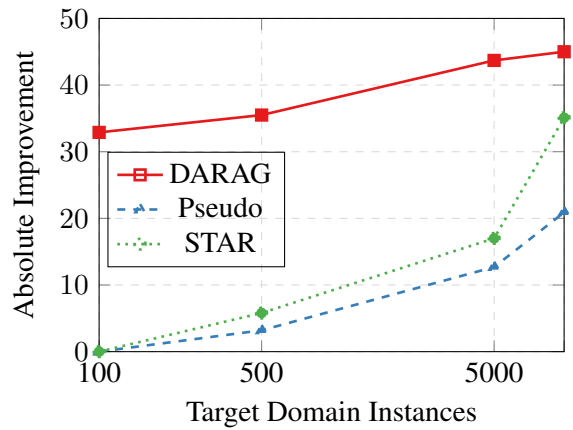
\begin{figure}[!t]
    \centering
    \begin{tikzpicture}
    \begin{axis}[
        xlabel={Target Domain Instances},
        ylabel={Absolute Improvement},
        xmin=100, xmax=10000,
        ymin=0, ymax=50,
        xtick={100, 500, 5000, 10000},
        xticklabels={100, 500, 5000},
        ytick={0, 10, 20, 30, 40, 50},
        xmode=log,
        log basis x={10},
        legend style={at={(0.4,0.6)}, fill=white, fill opacity=0.0, draw opacity=1, text opacity=1}, 
        width=1.0\linewidth,
        height=6cm,
        grid=major,
        grid style={dashed, gray!30},
        tick align=inside
    ]
    \definecolor{Method1Color}{RGB}{228,26,28} 
    \definecolor{Method2Color}{RGB}{55,126,184} 
    \definecolor{Method3Color}{RGB}{77,175,74} 

    \addplot[
        color=Method1Color,
        mark=square*,
        line width=1pt
    ]
    coordinates {
        (100, 32.9) (500, 35.5) (5000, 43.7) (10000, 45.0)
    };
    \addlegendentry{DARAG}

    \addplot[
        color=Method2Color,
        mark=triangle*,
        line width=1pt,
        dashed
    ]
    coordinates {
        (100, 0) (500, 3.2) (5000, 12.7) (10000, 20.9)
    };
    \addlegendentry{Pseudo}

    \addplot[
        color=Method3Color,
        mark=*,
        line width=1pt,
        dotted
    ]
    coordinates {
        (100, 0) (500, 5.8) (5000, 17.0) (10000, 35.1)
    };
    \addlegendentry{STAR}
    \end{axis}
    \end{tikzpicture}
    \vspace{-2mm}
    \caption{\small Comparison of DARAG with other methods on low-resource source-free UDA (LS $\rightarrow$ Vox). DARAG outperforms other methods with significant improvements.}
    \label{fig:abs_improvement}
    \vspace{-4mm}
\end{figure}

\section{Conclusion}
\vspace{-1mm}
We present DARAG, a novel approach for enhancing GEC in ASR systems. Our study reveals that GEC models struggle to generalize due to the limited diversity of error types encountered during training. To address this, we introduce two key improvements: (i) a synthetic data augmentation strategy that generates hypotheses with diverse errors resembling those the ASR model might realistically produce on a test set, and (ii) a retrieval-augmented NE correction mechanism. DARAG achieves performance superior to our compared methods.

\section*{Limitations}
As part of future work, we would like to work on the following limitations of our proposed DARAG approach:

\begin{enumerate}
    \item when the NE database is large, semantic similarity may result in the retrieval of multiple phonetically similar named entities, potentially causing confusion for the GEC model in choosing the correct entity. To address this, we plan to develop phoneme-aware NE retrieval methods to enhance retrieval accuracy.

    \item The use of synthetic data generated by LLMs could introduce biases inherent to the language models, potentially affecting the GEC model's performance. In future work, we aim to explore strategies for mitigating such biases to ensure more robust error correction.

    \item Although DARAG involves additional computational overhead for generating synthetic data, we anticipate that as model efficiency improves and lighter architectures become available, the overhead will be reduced, leading to even greater gains in performance.

    \item We only study ASR datasets in the English language. Future work includes evaluating DARAG's performance in low-resource languages beyond English.

\end{enumerate}

\bibliography{anthology,custom}

\appendix

\section{Appendix}
\label{sec:appendix}
In the Appendix, we provide:

\begin{enumerate}
\setlength\parskip{0em}
\item Section~\ref{sec:prompts}: Prompts
\item Section~\ref{sec:llm_copying}: Analysis of LLMs copying the training data 
\item Section~\ref{sec:darag_wo_voiceclone}: Performance of DARAG w/o Voice Cloning 
\item Section~\ref{sec:id_ood}: In-Domain Performance After Out-of-Domain Adaptation 
\item Section~\ref{sec:hyper-parameter}: Key Hyper-Parameter Tuning Results 
\item Section~\ref{sec:generation_examples}: Examples of DARAG Generations 
\item Section~\ref{sec:exam_correc}: Examples of DARAG Corrections 
\item Section~\ref{sec:additional}: Additional Details 

\end{enumerate}

\section{Prompts}
\label{sec:prompts}

We prompt LLaMa-2-Instruct in batched-mode with a temperature of 0.7 and top-$p$ of 1. We use this setting throughout all our experiments for generation and correction. We use the below prompt to generate synthetic transcripts using LLaMa-2-Instruct:
\vspace{30mm}
\begin{comparisonbox}
You need to act as a synthetic data generator. I will provide you with some example transcripts from a speech recognition dataset that I have transcribed using an ASR model. The transcripts are not related to each other. You need to first understand the nature of the spoken utterances from the transcripts and analyze their distinct features, like domain, style, length, etc. Next, with what you understood, you need to generate 2 short and diverse utterances with the same properties but diverse content. Each utterance should be a single sentence. Please include named entities as and when possible, but it is not necessary. Keep the utterances short and in line with the examples. Your generated transcripts should be coherent. Here are the example transcripts, one in each line:\{\}. Return a JSON with 2 keys named "First Transcript" and "Second Transcript" with the values as the generated transcripts.
\end{comparisonbox}

\section{Are LLMs Just Replicating the Original Training Data?}
\label{sec:llm_copying}

Previous research has suggested that LLMs may memorize open-domain ASR training transcripts~\cite{liu2024recent,team2023gemini}, raising the risk of replicating training data while genrating synthetic data. To evaluate whether this occurs with DARAG, we perform two checks: (i) We use SentenceBERT to calculate the cosine similarity between each generated transcript and all transcripts in the original training set, reporting the average semantic similarity across instances in Table~\ref{tab:similarity} ii) We compute the BLEU score for each generated transcript, using the transcript with the highest cosine similarity from the previous step as a reference. Table~\ref{tab:similarity} shows the average BLEU scores across BLEU\textsubscript{1}, BLEU\textsubscript{2}, and BLEU\textsubscript{3}. The low BLEU scores indicate that DARAG does not simply replicate the training data. The semantic similarity indicates that with DARAG generates transcripts that are consistent with the domain.

\begin{table}[h]
    \centering
    \begin{tabular}{lcc}
        \toprule
        \textbf{Dataset} & \textbf{Similarity} & \textbf{BLEU} \\
        \midrule
        LS & 0.32 & 0.12 \\
        Vox & 0.29 & 0.10 \\
        SPGI & 0.25 & 0.06 \\
        Giga & 0.22 & 0.13 \\
        TED & 0.26 & 0.14 \\
        \bottomrule
    \end{tabular}
    \vspace{-2mm}
    \caption{\small Semantic similarity and BLEU scores between original and generated transcripts across all datasets.}
    \label{tab:similarity}
\end{table}

\section{DARAG w/o Voice Cloning}
\label{sec:darag_wo_voiceclone}

Table~\ref{tab:wo_voice_clone} compares the performance of DARAG in both ID and OOD scenarios, with and without voice cloning. As discussed in Section~\ref{subsec:synthetic}, voice cloning via TTS allows the model to generate synthetic speech that, when transcribed, produces hypotheses containing errors similar to those encountered during testing in that domain. As shown in the table, DARAG experiences a performance drop without voice cloning, with a more significant decline in OOD scenarios.

\begin{table}[h]
\centering
\small
\resizebox{0.48\textwidth}{!}{
\begin{tabular}{cllc}
\toprule
\textbf{Test} & \textbf{Method}  & \textbf{Train} & \textbf{WER ($\downarrow$)} \\
\midrule
\multirow{6}{*}{Vox}
&\myccfour Baseline &\myccfour Vox &\myccfour 10.1\\
& \hspace{0.25cm}+DARAG & Vox & 8.6 \\ 
& \hspace{0.25cm}+DARAG w/o Voice Cloning & Vox & 8.8\\
&\myccfour Baseline &\myccfour LS &\myccfour 14.9\\
& \hspace{0.25cm}+DARAG & LS & 10.0 \\ 
& \hspace{0.25cm}+DARAG w/o Voice Cloning & LS & 12.2\\

\cmidrule(l){1-4}
\multirow{6}{*}{LS}
&\myccfour Baseline &\myccfour LS &\myccfour 8.4\\
\multirow{6}{*}{(Other)}& \hspace{0.25cm}+DARAG & LS & 6.4 \\ 
& \hspace{0.25cm}+DARAG w/o Voice Cloning & LS & 7.3\\
&\myccfour Baseline &\myccfour Vox &\myccfour 13.7\\
& \hspace{0.25cm}+DARAG & Vox & 11.9 \\ 
& \hspace{0.25cm}+DARAG w/o Voice Cloning & Vox & 14.5\\
\bottomrule
\end{tabular}
}
\caption{\small Performance comparison of DARAG with and without voice cloning. Performance drops sharply without voice cloning, especially in OOD scenrios, thereby confirming the importance of the voice cloning for generating augmentations.}
\label{tab:wo_voice_clone}
\end{table}

\section{In-Domain Performance in Out-of-Domain Settings}
\label{sec:id_ood}

Table~\ref{tab:id_ood} presents the performance of DARAG on in-domain tests after augmenting the hypotheses dataset with OOD hypotheses-transcription pairs. The results demonstrate that DARAG maintains its performance on the in-domain test with only a negligible drop.

\begin{table}[h]
\centering
\small
\resizebox{0.48\textwidth}{!}{
\begin{tabular}{clccc}
\toprule
\textbf{Test} & \textbf{Method}  & \textbf{Train} & \textbf{OOD Adapt.} & \textbf{WER ($\downarrow$)} \\
\midrule
\multirow{5}{*}{Vox}
&\myccfour Baseline &\myccfour -   &\myccfour -   &\myccfour 10.1\\
&\myccfour \hspace{0.25cm}+DARAG &\myccfour Vox &\myccfour - &\myccfour 8.6 \\
& \hspace{0.25cm}+DARAG & Vox & LS & 8.9 \\ 
& \hspace{0.25cm}+DARAG & Vox & SPGI  & 9.0\\
& \hspace{0.25cm}+DARAG & Vox & TED & 9.0 \\\cmidrule(l){1-5}
\multirow{5}{*}{LS}
&\myccfour Baseline &\myccfour -   &\myccfour -   &\myccfour 8.4\\
\multirow{5}{*}{(Other)}&\myccfour \hspace{0.25cm}+DARAG &\myccfour LS &\myccfour - &\myccfour 6.4 \\
& \hspace{0.25cm}+DARAG & LS & Vox & 7.5 \\ 
& \hspace{0.25cm}+DARAG & LS & SPGI  & 7.8\\
& \hspace{0.25cm}+DARAG & LS & TED & 6.9\\ 
\bottomrule
\end{tabular}
}
\caption{\small Performance comparison of DARAG across different settings. OOD Adapt. refers to the dataset for which synthetic data was generated and augmented to the original hypotheses for GEC training. Our results show that, even with the addition of synthetically generated training data, DARAG maintains its in-domain performance. Furthermore, improvements in a specific domain occur only when the augmentations are consistent with that domain. This approach ensures that the errors used for training match the characteristics of those the ASR model will encounter during testing.}
\label{tab:id_ood}
\end{table}

\section{Hyper-parameter Tuning}
\label{sec:hyper-parameter}

\subsection{Effect of $k$ for NE retrieval}
Table~\ref{tab:k_value} compares the performance of DARAG across various values of $k$ for NE retrieval. We choose two in-domain settings as our main experiments show NE retrieval is most effective in in-domain scenarios. We show both higher and lower values of $k$ can lead to a drop in performance and find 5 as the most optimal value. Higher values of $k$ can retrieve irrelevant NEs and confuse the GEC model. Lower values of $k$ can lead to cases where the GT NE is not retrieved.

\begin{table}[h]
\centering
\small
\resizebox{0.48\textwidth}{!}{
\begin{tabular}{cccccc}
\toprule
\textbf{Test} & \textbf{$k$=1} & \textbf{$k$=2} & \textbf{$k$=5} & \textbf{$k$=7} & \textbf{$k$=9} \\
\midrule
Vox & 87.8 & \underline{88.7} & \textbf{90.0} & 87.9 & \underline{87.8} \\
LS (Other)  & 94.5 & \underline{94.5} & \textbf{96.4} & 93.9 & 93.3 \\
\bottomrule
\end{tabular}
}
\vspace{-2mm}
\caption{\small Performance comparison of DARAG on two in-domain settings with various values of $k$ for NE retrieval.}
\label{tab:k_value}
\end{table}

\subsection{Effect of $n_\text{small}$ in OOD settings}

Table~\ref{tab:ner_end_appendix} compares the performance of DARAG across various values of $n_\text{small}$. Larger $n_\text{small}$ can lead to more diverse and consistent augmentations, improving performance. For our primary experiments, we stick to 100 to keep our setting ultra-low-resource.

\begin{table}[h]
\centering
\small
\resizebox{0.35\textwidth}{!}{
\begin{tabular}{ccccc}
\toprule
\textbf{Test} & \textbf{10} & \textbf{50} & \textbf{100} & \textbf{500} \\
\midrule
Vox & 15.2 & 11.3 & \underline{10.0} & \textbf{9.5} \\
SPGI  & 17.9 & 14.1 & \underline{12.0} & \textbf{11.7} \\
\bottomrule
\end{tabular}
}

\caption{\small Performance comparison of DARAG on two OOD settings (with LS as training set) with various values of $n_\text{small}$. Larger values can lead to improved performance.}
\label{tab:ner_end_appendix}
\end{table}

\subsection{Effect of $n_\text{syn}$}

Table~\ref{tab:scaling} compares the performance of DARAG using different values of $n_\text{syn}$, represented as a factor of $n$ (the size of the original training set for the target dataset in an OOD setting). Increasing the number of synthetic samples (higher $n_\text{syn}$) can provide more diverse and consistent augmentations in OOD settings, resulting in better performance. However, the improvements plateau beyond a certain point. For our main experiments, we use $n_\text{syn}=1$ due to resource limitations.

\begin{table}[h]
\centering
\small
\resizebox{0.30\textwidth}{!}{
\begin{tabular}{lcccc}
\toprule
\textbf{Test} & \textbf{0.5$\times$} & \textbf{1$\times$} & \textbf{2$\times$} & \textbf{5$\times$}\\
\midrule
Vox   & 13.1 & 10.0 & \textbf{9.6} & \underline{9.7}\\
SPGI  & 14.2 & \underline{12.0} & \textbf{11.3} & \textbf{11.3}\\
\bottomrule
\end{tabular}
}
\caption{\small Performance comparison of DARAG on two OOD settings (with LS as training set) across different scaling factors of $n_\text{syn}$ relative to $n$. More synthetic samples can lead to improved performance, but plateaus beyond a certain point.}
\label{tab:scaling}
\end{table}

\section{Examples of Generated Transcripts}
\label{sec:generation_examples}

Table~\ref{tab:example_transcripts} provides examples of synthetically generated transcripts for each dataset from our evaluation setup. The transcripts are coherent and consistent with the characteristics of the domain.

\begin{table*}[t]
\centering
\small
\resizebox{\textwidth}{!}{
\begin{tabular}{p{2cm}|p{10cm}}
\toprule
\multicolumn{1}{l}{\textbf{Dataset}} & \multicolumn{1}{c}{\textbf{Synthetic Transcripts}} \\ \midrule
LibriSpeech & the duke entered the grand hall as the musicians began playing a lively gavotte \\
LibriSpeech & her highness attended the gala wearing the renowned emerald necklace from the royal collection \\ 
\midrule
SPGI & Sarah, can we reassess the projected growth for the third quarter and adjust our targets accordingly? \\

SPGI & Our current expectation is to maintain a minimum margin of 40\%, though market conditions may lead to some adjustments. \\ \midrule

GigaSpeech & please navigate to the settings page to update your api key and configure the callback url. \\
GigaSpeech & she served as the vice chair of the european data protection board for three years before joining the united nations privacy task force. \\ \midrule

VoxPopuli & as the smoke cleared the battered zeppelin drifted slowly back towards the enemy's encampment \\
VoxPopuli & yet i shall not yield to their demands but will defend my honor just as young frederick once did in times of great peril \\ \midrule

TED & we are often overwhelmed by too many options and that can make even simple decisions difficult to navigate \\
TED & i must admit that my journey has had its ups and downs but in the end i found exactly what i was looking for \\ \midrule

\end{tabular}}
\caption{\small Examples of generated transcripts by the DARAG methodology.}
\label{tab:example_transcripts}
\end{table*}

\section{Examples of DARAG Corrections}
\label{sec:exam_correc}

Table~\ref{tab:corrections} qualitatively compares DARAG with traditional GEC on various instances from benchmark datasets. We show that DARAG is able to accurately correct NEs which traditional GEC cannot. Additionally, DARAG shows superior performance in OOD scenarios.

\begin{table*}[t]
\centering
\small
\resizebox{\textwidth}{!}{
\begin{tabular}{p{2cm}|p{4cm}|p{4cm}|p{4cm}}
\toprule
\multicolumn{1}{l}{\textbf{Dataset}} & \multicolumn{1}{c}{\textbf{ASR Transcription}} & \multicolumn{1}{c}{\textbf{Traditional GEC}} & \multicolumn{1}{c}{\textbf{DARAG}}\\ \midrule
LibriSpeech Other & how \textcolor{red}{eye} wish you could get me a coffee of that pitcher phi\textcolor{red}{l}lip laura said in \textcolor{red}{treating lee} & how i wish you could get me a coffee of that pitcher phi\textcolor{red}{l}lip laura said in treatingly & how i wish you could get me a copy of that picture philip laura said treatingly \\ \midrule

LibriSpeech Other (OOD on Vox) & but she fixed \textcolor{red}{up on} a \textcolor{red}{pitcher} which she said she preferred \textcolor{red}{too} anything she had \textcolor{red}{scene} in the galley & but she fixed \textcolor{red}{up on} a \textcolor{red}{pitcher} which she said she preferred \textcolor{red}{too} anything she had \textcolor{red}{scene} in the galley & but she fixed upon a picture which she said she preferred to anything she had seen in the gallery \\ 
\midrule

SPGI & and we expect once the \textcolor{red}{Sharon} Nation Credit gets taken care of, we're in a arrange where we will be managing in flows and out flows on a normal & and we expect once the \textcolor{red}{Sharon} Nation Credit gets taken care of, we're in a arrange where we will be managing in flows and out flows on a normal & and we expect once the Shared National Credit gets taken care of, we're in a range where we will be managing inflows and outflows on a normal \\ \midrule

SPGI (OOD on Vox) & obviously, the confidence level on future and growing exploration in the \textcolor{red}{Golf} of Mexico, in South East Asia. So & obviously, the confidence level on future and growing exploration in the \textcolor{red}{Golf} of Mexico, in South East Asia. So  & obviously, the confidence level on future and growing exploration in the Gulf of Mexico, in Southeast Asia. So \\ \midrule

GigaSpeech & TRULY THE EIGHT WONDER OF THE WORLD \textcolor{red}{SEAN ELLIOT.} THANK YOU SO MUCH. & TRULY THE EIGHT WONDER OF THE WORLD \textcolor{red}{SEAN ELLIOT.} THANK YOU SO MUCH. & TRULY THE EIGHTH WONDER OF THE WORLD SHAWN ELLIOTT . THANK YOU SO MUCH . \\ \midrule
GigaSpeech (OOD on Vox) & MICROSOFT FIRED BACK WITH ITS OWN SEARCH \textcolor{red}{INJUN} & MICROSOFT FIRED BACK WITH ITS OWN SEARCH \textcolor{red}{INJUN} & MICROSOFT FIRED BACK WITH ITS OWN SEARCH ENGINE \\ \midrule

VoxPopuli & we need mores sources we need mores pipes than one from \textcolor{red}{rush ya} & we need mores sources we need mores pipes than one from \textcolor{red}{rush ya} & we need more sources we need more pipes than one from russia \\ \midrule
VoxPopuli (OOD on LibriSpeech) & may \textcolor{red}{i in decay} however that the protection of arbitration agreements should not limited the free circulation of judgments in the union  & may i indicate however that the protection of arbitration agreements should not limited the free circulation of judgments in the union & may i indicate however that the protection of arbitration agreements should not limit the free circulation of judgements in the union \\ \bottomrule

\end{tabular}}
\caption{\small Examples of incorrect ASR transcriptions and their corresponding corrections by DARAG.}
\label{tab:corrections}
\end{table*}

\section{Additional Details}
\label{sec:additional}

{\noindent{\textbf{Compute details.}}} For all our pre-training and fine-tuning experiments, we used four NVIDIA A6000-48GB GPUs. Each training requires 4-24 hours. 
\vspace{0.5mm}

{\noindent{\textbf{Potential Risk.}}} As mentioned in the limitations section of the paper, DARAG might encode biases inherent to the LLM. This might lead to unsafe generations and corrections. Additionally, voice cloning systems used as part of our method can be employed to create deep fake voices.

{\noindent \textbf{Software and Packages details.}} We implement all our models in PyTorch ~\footnote{\url{https://pytorch.org/}} and use Parler-TTS~\footnote{\url{https://github.com/huggingface/parler-tts}} and LLaMa-2~\footnote{\url{https://huggingface.co/meta-llama}}. We employ ESPnet~\cite{watanabe2018espnet} for training our ASR models.

{\noindent{\textbf{Use of AI models.}}} We used GPT-4 for rephrasing certain parts of the writing.

{\noindent{\textbf{Datasets.}}} Dataset details, together with statistics are provided below:

{\noindent \textbf{LibriSpeech}~\footnote{https://www.openslr.org/12}} The LibriSpeech dataset is a large-scale corpus of approximately 1,000 hours of 16kHz English speech derived from audiobooks in the LibriVox project, with text sourced primarily from Project Gutenberg. It is split into training sets (100hr, 360hr, and 500hr) and dev/test sets categorized as dev clean(5hr), dev other(5hr), test clean(5hr), and test other(5hr) based on transcription difficulty. The dataset also includes n-gram language models and texts with 803 million tokens and 977,000 unique words, making it valuable for Automatic Speech Recognition (ASR) research. 

{\noindent \textbf{SPGISpeech}~\footnote{https://datasets.kensho.com/datasets/spgispeech}} SPGISpeech is a large-scale speech transcription dataset containing 5,000 hours of professionally transcribed financial audio, including company earnings calls with a variety of L1 and L2 English accents. It features approximately 50,000 speakers and offers high-quality transcripts that have been thoroughly edited for accuracy, including proper punctuation, capitalization, and denormalization of non-standard words. The audio is split into 5 to 15-second slices, formatted as single-channel, 16kHz, 16-bit WAV files, making it ideal for training advanced speech recognition models.

{\noindent \textbf{VoxPopuli}~\footnote{https://github.com/facebookresearch/voxpopuli}} VoxPopuli is a large-scale multilingual speech corpus designed for tasks like representation learning, semi-supervised learning, and interpretation. It offers 400,000 hours of unlabeled speech in 23 languages, resulting in 8K-24K hours of data for each language, 1,800 hours of transcribed speech in 16 languages, and 17,300 hours of speech-to-speech interpretation across 15 language pairs. In transcribed speech, the filtered utterances are split into train, development and test sets with disjoint speakers and target duration ratio (18:1:1). Additionally, it includes 29 hours of transcribed non-native English speech for research on accented speech in ASR.

{\noindent \textbf{GigaSpeech}~\footnote{https://github.com/SpeechColab/GigaSpeech}} 
GigaSpeech is a large-scale English speech recognition corpus with 10,000 hours of training set of high-quality human-transcribed audio for supervised learning, 12 hours of dev set, and 40 hours of test set. It is designed for both supervised and unsupervised/semi-supervised learning tasks, covering a wide range of domains. It is particularly suited for large-scale speech recognition model training and adaptation.

{\noindent \textbf{TED-LIUM (v1)}~\footnote{https://www.openslr.org/7/}} 
The TED-LIUM corpus is a dataset of English-language TED talks, featuring transcriptions of talks sampled at 16kHz. It contains approximately 118 to 452 hours of transcribed speech data, with 56,803 examples in the training set, 1,469 in the test set, and 591 in the validation set. This dataset is widely used for Automatic Speech Recognition (ASR) research and model training.

All datasets used in our paper are openly available for download and free to use to academic research.

\end{document}